\documentclass[aps,prl,preprint,superscriptaddress,floatfix,amsmath,amsfonts,showpacs]{revtex4}

\usepackage{bm,graphicx}

\begin{document}


\title{Evidence of a Pionic Enhancement Observed in 
${}^{16}{\rm O}(p,p'){}^{16}{\rm O}(0^-,T=1)$ at 295 MeV}


\author{T.~Wakasa}
\affiliation{Department of Physics, Kyushu University,
Fukuoka 812-8581, Japan}
\author{G.~P.~A.~Berg}
\affiliation{Kernfysisch Versneller Instituut, 
Zernikelaan 25, 9747 AA Groningen, The Netherlands}
\author{H.~Fujimura}
\affiliation{Department of Physics, Kyoto University,
Kyoto 606-8502, Japan}
\author{K.~Fujita}
\affiliation{Research Center for Nuclear Physics, Osaka University,
Osaka 567-0047, Japan}
\author{K.~Hatanaka}
\affiliation{Research Center for Nuclear Physics, Osaka University,
Osaka 567-0047, Japan}
\author{M.~Ichimura}
\affiliation{Faculty of Computer and Information Sciences, Hosei University, 
Tokyo 184-8584, Japan}
\author{M.~Itoh}
\affiliation{Research Center for Nuclear Physics, Osaka University,
Osaka 567-0047, Japan}
\author{J.~Kamiya}
\affiliation{Accelerator Group, Japan Atomic Energy Research Institute,
Ibaraki 319-1195, Japan}
\author{T.~Kawabata}
\affiliation{Center for Nuclear Study, The University of Tokyo,
Tokyo 113-0033, Japan}
\author{Y.~Kitamura}
\affiliation{Research Center for Nuclear Physics, Osaka University,
Osaka 567-0047, Japan}
\author{E.~Obayashi}
\affiliation{Research Center for Nuclear Physics, Osaka University,
Osaka 567-0047, Japan}
\author{H.~Sakaguchi}
\affiliation{Department of Physics, Kyoto University,
Kyoto 606-8502, Japan}
\author{N.~Sakamoto}
\affiliation{Research Center for Nuclear Physics, Osaka University,
Osaka 567-0047, Japan}
\author{Y.~Sakemi}
\affiliation{Research Center for Nuclear Physics, Osaka University,
Osaka 567-0047, Japan}
\author{Y.~Shimizu}
\affiliation{Research Center for Nuclear Physics, Osaka University,
Osaka 567-0047, Japan}
\author{H.~Takeda}
\affiliation{The Institute of Physical and Chemical Research,
Saitama 351-0198, Japan}
\author{M.~Uchida}
\affiliation{Department of Physics, Tokyo Institute of Technology,
Tokyo 152-8550, Japan}
\author{Y.~Yasuda}
\affiliation{Department of Physics, Kyoto University,
Kyoto 606-8502, Japan}
\author{H.~P.~Yoshida}
\affiliation{Research and Development Center for Higher Education, 
Kyushu University,
Fukuoka 810-8560, Japan}
\author{M.~Yosoi}
\affiliation{Department of Physics, Kyoto University,
Kyoto 606-8502, Japan}

\date{\today}

\begin{abstract}
 The cross section of the 
${}^{16}{\rm O}(p,p'){}^{16}{\rm O}(0^-,T=1)$ scattering 
was measured at a bombarding energy of 295 MeV 
in the momentum transfer range of 
$1.0\,\mathrm{fm^{-1}}$ $\le$ $q_{\rm c.m.}$ $\le$ $2.1\,\mathrm{fm^{-1}}$.
 The isovector $0^-$ state at $E_x$ = 12.8 MeV is clearly 
separated from its neighboring states owing to the high 
energy resolution of about 30 keV.
 The cross section data were compared with distorted wave impulse 
approximation (DWIA) calculations employing shell-model 
wave functions.
 The observed cross sections around 
$q_{\rm c.m.}$ $\simeq$ $1.7\,{\rm fm^{-1}}$ 
are significantly larger than predicted by these 
calculations, suggesting pionic enhancement 
as a precursor of pion condensation in nuclei.
 The data are successfully reproduced by 
DWIA calculations using random phase approximation
response functions including the $\Delta$ isobar 
that predict pionic enhancement.
\end{abstract}


\maketitle


 The search for pionic enhancements
in nuclei has a long and interesting history.
 These phenomena can be considered as a precursor
of the pion condensation \cite{migdal} 
that would be realized in neutron stars.
 Enhancements of the {\it M}1 cross section in  
proton inelastic scattering 
\cite{prl_42_1034_1979,plb_89_327_1980,plb_91_328_1980,plb_92_265_1980} 
and of the ratio $R_L/R_T$, 
the spin-longitudinal (pionic) response function $R_L$ to the 
spin-transverse one $R_T$, 
in the quasielastic scattering (QES) region 
\cite{plb_92_153_1980,npa_379_429_1982} were expected 
around a momentum transfer 
$q_{\rm c.m.}$ $\simeq$ $1.7\,{\rm fm^{-1}}$.
 However, the experimental data did not reveal any
pionic enhancements 
\cite{prc_23_1858_1981,prl_73_3516_1994,prc_59_3177_1999}.
 Several explanations exist to answer the question why
no pionic enhancements were observed.
 For example, Bertsch, Frankfurt, and Strikman 
\cite{science_259_773_1993} 
suggest the modification of gluon properties 
in the nucleus that suppresses the pion field.
 Brown {\it et al.} \cite{npa_593_295_1995} 
suggest the partial restoration of 
chiral invariance with density. 
 However, we should note that the {\it M}1 cross section 
includes both pionic and non-pionic contributions and 
$R_L/R_T$ is the ratio to the non-pionic $R_T$.
 Thus, in these indirect measurements, 
the pionic enhancement might be masked by the 
contribution from the non-pionic component.
 Recent analyses of the QES data 
\cite{prc_69_054609_2004,nucl_ex_0411055} 
show a pionic enhancement in the spin-longitudinal 
cross section that well represents the $R_L$, 
and suggest that the lack of enhancements of $R_L/R_T$ 
is due to the non-pionic component.
 
 In order to measure the pionic enhancement directly,
it is desirable to investigate isovector $J^{\pi}$ = $0^-$, 
$0^{\pm}$ $\rightarrow$ $0^{\mp}$ excitations 
because they carry the same quantum numbers as the pion and 
they are free from non-pionic contributions.
 Orihara {\it et al.} \cite{prl_49_1318_1982} 
measured the angular distribution of the 
${}^{16}{\rm O}(p,n){}^{16}{\rm N}(0^-,$ 0.12 MeV) reaction 
at $T_p$ = 35 MeV.
 They reported discrepancies between distorted wave Born approximation
calculations and their data in the range of
$q_{\rm c.m.}$ = 1.4--2.0 ${\rm fm^{-1}}$ that 
might be a signature of pionic enhancement.
 However, in the proton inelastic scattering to the $0^-$, 
$T=1$ state in ${}^{16}{\rm O}$ at $T_p$ = 65 MeV,
such an enhancement was not observed \cite{prc_30_746_1984}.
 The differences between these $(p,n)$ and $(p,p')$ results might 
indicate contributions from complicated reaction mechanisms 
at these low incident energies.
 To our knowledge, there are no published experimental data 
for the $0^-$, $T$ = 1 state 
at intermediate energies of $T_p$ $>$ 100 MeV where 
reaction mechanisms are expected to be simple.

 In this Letter, we present the measurement of the cross section
for the excitation of the 
$0^-$, $T$ = 1 state at $E_x$ = 12.8 MeV in ${}^{16}{\rm O}$ 
using inelastic proton scattering at 295 MeV incident energy.
 The results are compared with distorted wave impulse 
approximation (DWIA) calculations with shell-model (SM) 
wave functions.
 Evidence of a pionic enhancement is clearly observed 
from a comparison between experimental and theoretical results.
 The data are also compared with DWIA calculations employing 
random phase approximation (RPA) response functions 
including the $\Delta$ isobar 
in order to assess the pionic enhancement quantitatively. 

 The measurement was carried out by using the 
West-South Beam Line (WS-BL) \cite{nim_a482_79_2002} 
and the Grand Raiden (GR) spectrometer \cite{nim_a422_484_1999}
at the Research Center for Nuclear Physics, Osaka University.
 The WS-BL provides the beam with 
lateral and angular dispersions of 37.1 m and $-20.0$ rad, 
respectively, which 
satisfy the dispersion matching conditions for GR.
 The beam bombarded a windowless and self-supporting 
ice (${\rm H_2O}$) target \cite{nim_a459_171_2001} 
with a thickness of 14.1 ${\rm mg/cm^2}$.
 Protons scattered from the target were momentum analyzed 
by the high-resolution GR spectrometer 
with a typical resolution of $\sim$30 keV FWHM.
 The beam energy was determined to be $295\pm 1$ MeV 
by using the kinematic energy shift between elastic 
scattering from ${}^{1}{\rm H}$ and ${}^{16}{\rm O}$.
 The yields of the scattered protons were extracted using the 
peak-shape fitting program {\sc allfit} \cite{allfit}.

\begin{figure}
\includegraphics[width=0.9\linewidth,clip]{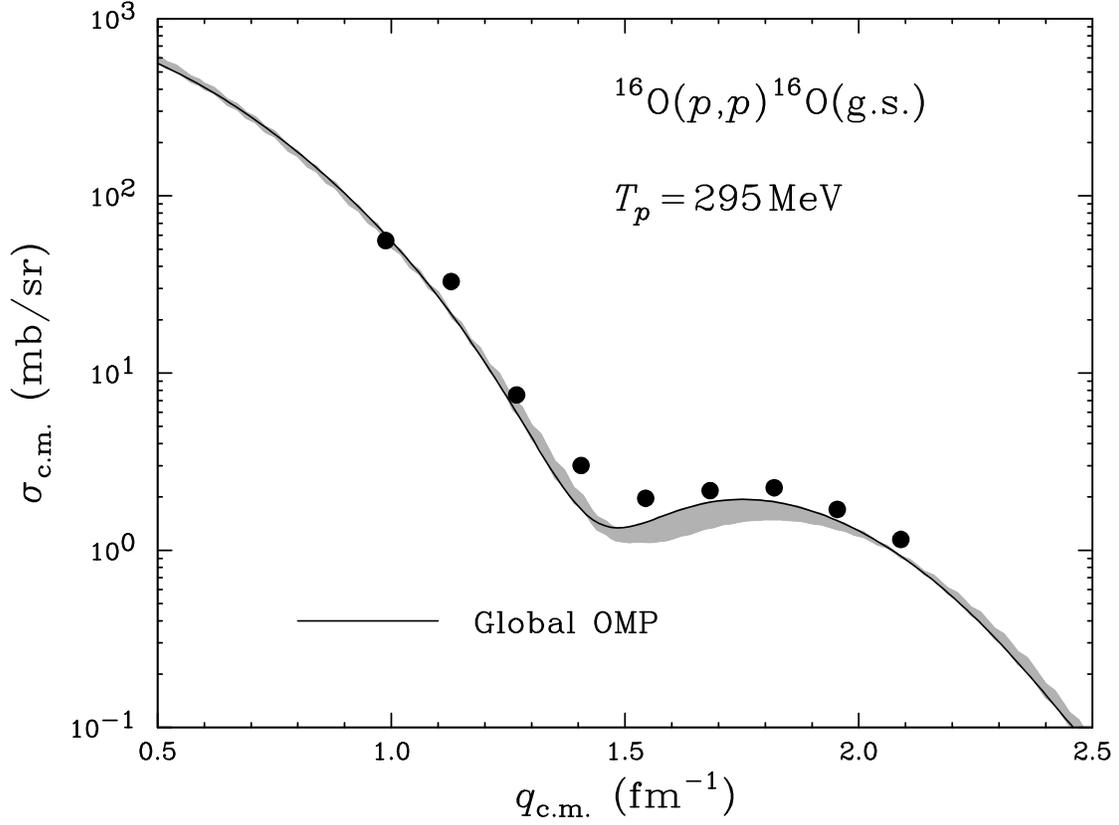}%
\caption{
 The measurement of the cross section for 
${}^{16}{\rm O}(p,p)$ at $T_p$ = 295 MeV.
 The solid curve is the theoretical prediction using the 
global OMP for ${}^{16}{\rm O}$.
 The band represents the ambiguity of the prediction 
as explained in the text.
\label{fig:elastic}}
\end{figure}

 The elastic scattering data on ${}^{16}{\rm O}$ are 
shown in Fig.~\ref{fig:elastic}.
 Differential cross sections were normalized to the known 
$p+p$ cross section \cite{said} 
by utilizing the data of protons scattered from 
the hydrogen present in the ice target.
 The data were analyzed using 
optical model potentials (OMPs) generated phenomenologically.
 The solid curve in Fig.~\ref{fig:elastic} is the result 
using the global OMP optimized for 
${}^{16}{\rm O}$ \cite{prc_41_2737_1990}.
 The band represents the results by using several
OMPs parametrized for nuclei from ${}^{12}{\rm C}$ 
to ${}^{208}{\rm Pb}$ with a smooth mass number 
dependence \cite{prc_41_2737_1990} that shows the ambiguity of 
the OMP for ${}^{16}{\rm O}$.
 The global OMP for ${}^{16}{\rm O}$ reproduces the 
experimental data reasonably well.
 The systematic uncertainty for the cross section 
normalization is estimated to be less than $\sim$10\% 
from this result.
 In the following, we will use this OMP
in DWIA calculations for inelastic scattering.

\begin{figure}
\includegraphics[width=0.9\linewidth,clip]{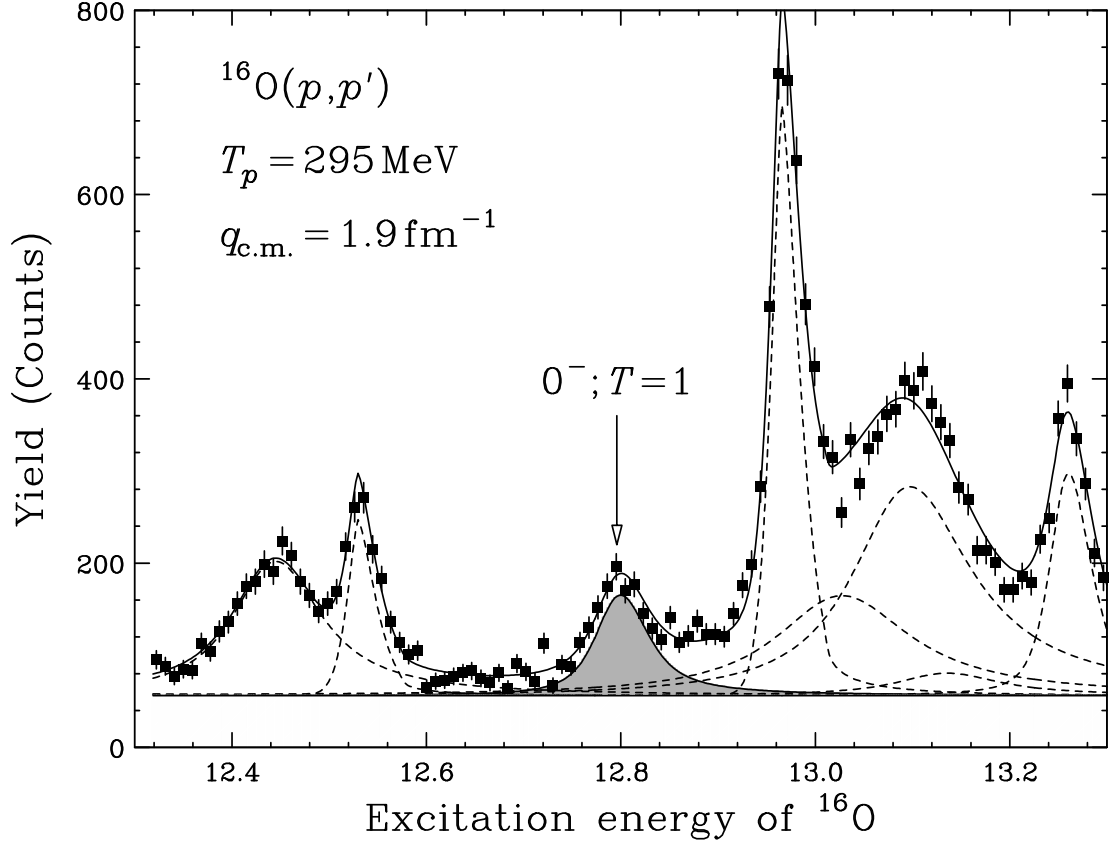}%
\caption{
 The excitation energy spectrum for 
${}^{16}{\rm O}(p,p')$ at 
$T_p$ = 295 MeV and $q_{\rm c.m.}$ = $1.9\,\mathrm{fm^{-1}}$.
 The curves show the reproduction of this spectrum 
with hyper-Gaussian and Lorentzian peaks and a continuum. 
\label{fig:fit}}
\end{figure}

 Figure~\ref{fig:fit} shows the excitation energy 
spectrum of the ${}^{16}{\rm O}(p,p')$ scattering at 
$q_{\rm c.m.}$ = $1.9\,\mathrm{fm}^{-1}$.
 The isovector $0^-$ state at $E_x$ = 12.8 MeV is clearly 
resolved from the neighboring states.
 The dashed curves represent the fits to the individual 
peaks while the straight line and solid curve 
represent the background and the sum of the peak 
fitting, respectively.
 Narrow peaks of ${}^{16}{\rm O}$ were described by a 
standard hyper-Gaussian line shape, and the peaks with intrinsic 
widths were described as Lorentzian shapes convoluted with a 
resolution function based on the narrow peaks.
 The positions and widths were taken from Ref.~\cite{npa_375_1_1982}.

\begin{figure}
\includegraphics[width=0.9\linewidth,clip]{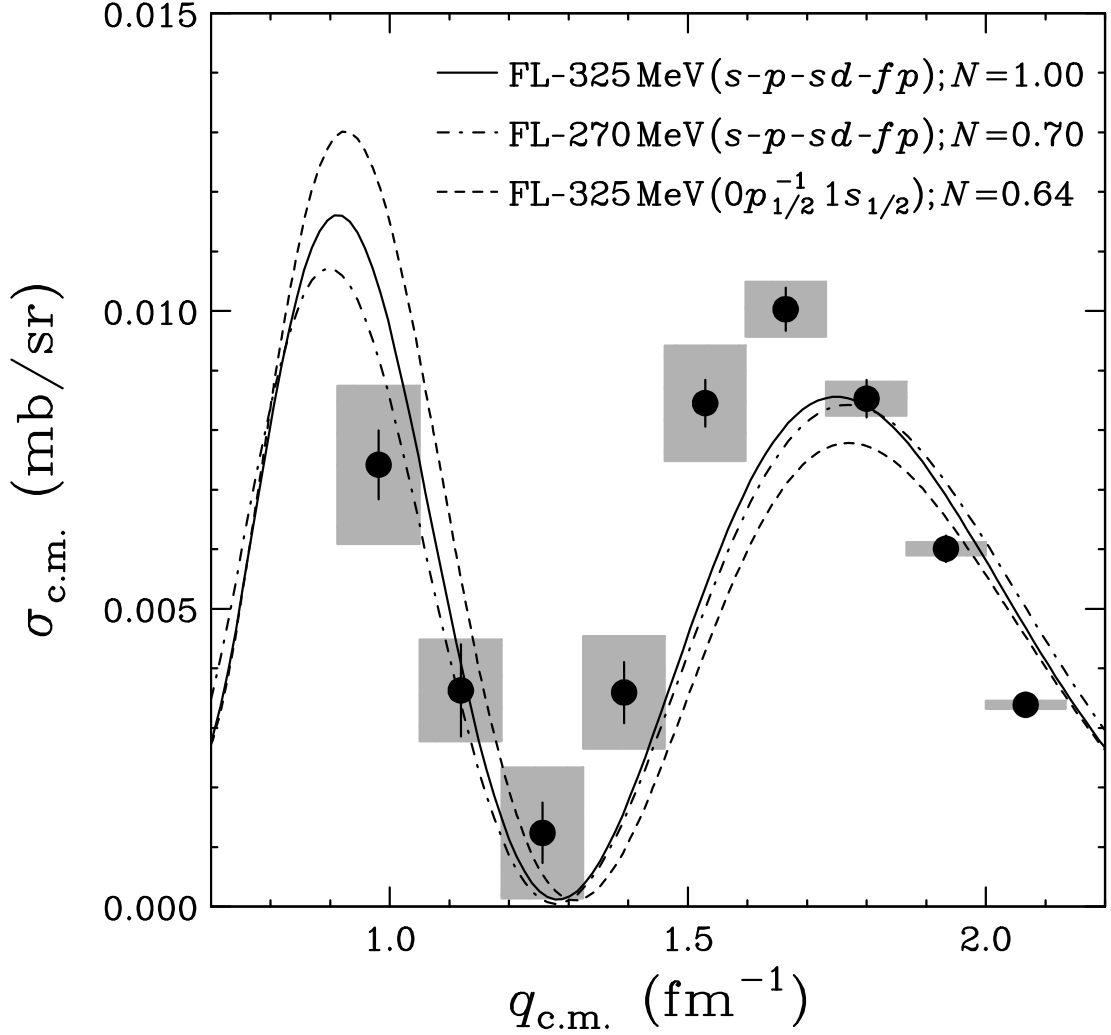}%
\caption{
 The measurement of the cross section of
${}^{16}{\rm O}(p,p'){}^{16}{\rm O}(0^-,T=1)$ at  $T_p$ = 295 MeV.
 The shaded areas represent the systematic uncertainties of the data.
 The solid (dash-dotted) curve is the DWIA result with 
the $t$-matrix parametrized at 325 (270) MeV employing 
the SM wave function in the $0s$-$0p$-$1s0d$-$1p0f$ model 
space.
 The dashed curve denotes the DWIA result with the
$t$-matrix at 325 MeV employing the pure 
$0p_{1/2}1s_{1/2}$ SM wave function.
\label{fig:xsec}}
\end{figure}

 Figure~\ref{fig:xsec} shows the measured data points and the
calculated curves of the cross sections of the 
$0^-$, $T=1$ transition in ${}^{16}{\rm O}(p,p')$ 
as a function of the momentum transfer $q_{\rm c.m.}$.
 The angular distribution was measured in the range 
of $q_{\rm c.m.}$ $\simeq$ $1.0\,\mathrm{fm^{-1}}$ 
to $\simeq$ $2.1\,\mathrm{fm^{-1}}$ 
starting near the second maximum at 
$q_{\rm c.m.}$ $\simeq$ $0.9\,\mathrm{fm^{-1}}$ and 
extending beyond the third maximum at 
$q_{\rm c.m.}$ $\simeq$ $1.7\,\mathrm{fm^{-1}}$.
 The error bars of the data points are the fitting uncertainties 
originating from the statistical uncertainties.
 The shaded areas represent the systematic uncertainties 
including the background subtraction.

 We performed DWIA calculations by using 
the computer code {\sc dwba98} \cite{dwba98}.
 The one-body density matrix elements (OBDME) for the isovector 
$0^-$ transition of ${}^{16}{\rm O}(p,p')$ were 
obtained from Ref.~\cite{prc_65_024322_2002}.
 This SM calculation was performed
in the $0s$-$0p$-$1s0d$-$0f1p$
configuration space by using phenomenological effective interactions.
 In the calculation, the ground state of ${}^{16}{\rm O}$ was 
described as a mixture of $0\hbar \omega$ (closed-shell) and 
$2\hbar\omega$ configurations.
 The single particle radial wave functions were generated 
by using a Woods-Saxon (WS) potential \cite{bohr_mottelson}, 
the depth of which was adjusted to 
reproduce the separation energies of the $0p_{1/2}$ orbits.
 The unbound single particle states were 
assumed to have a very small binding energy of 0.01 MeV 
to simplify the calculations.
 The {\it NN} {\it t}-matrix parametrized 
by Franey and Love \cite{prc_31_488_1985} at 325 MeV was used.
 The DWIA result is shown as the solid curve in Fig.~\ref{fig:xsec}.
 The calculation reproduces the data in the lower-$q_{\rm c.m.}$ region 
reasonably well, but they significantly underestimate the data
in the higher-$q_{\rm c.m.}$ region.
 Also, the data has a maximum at 
$q_{\rm c.m.}$ $\simeq$ $1.7\,\mathrm{fm^{-1}}$, 
whereas the maximum of the theoretical curve 
is slightly higher at 
$q_{\rm c.m.}$ $\simeq$ $1.8\,\mathrm{fm^{-1}}$.

 We investigated the sensitivity of the DWIA calculations 
to changes of the parameters involved.
 The dash-dotted curve represents the DWIA calculation 
with a different 
{\it t}-matrix parametrized at 270 MeV. 
 The result is systematically larger compared to the 
calculation with 
the {\it t}-matrix at 325 MeV.
 The dash-dotted curve is, therefore, multiplied by 
a factor of 0.7.
 The dashed curve denotes the calculation 
employing a different 
OBDME with a pure $0p_{1/2}^{-1}1s_{1/2}$ transition from the 
$0\hbar\omega$ (closed-shell) ground state.
 Auerbach and Brown \cite{prc_65_024322_2002} 
suggest that this isovector strength is 
quenched and spread by a $2\hbar\omega$ admixture.
 They obtained a quenching factor of $\sim$0.64.
 Thus we have applied this factor as a normalization factor 
to the result.
 We also performed a DWIA calculation with the 
radial wave functions generated with a harmonic oscillator potential
with a size parameter of $\alpha$ = $0.588\,\mathrm{fm^{-1}}$ 
\cite{prc_65_064316_2002}.
 The result is systematically larger compared to the 
calculation with the WS potential.
 However, their shapes of the angular distribution 
are very similar to each other.
 From these calculations we found that the shape
of the angular distribution 
is insensitive to changes of the input parameters.
 Thus it is difficult to understand the discrepancies 
between experimental and theoretical results 
within the framework of the standard DWIA employing SM wave functions.
 Therefore, in the following, we investigate
non-locality, RPA correlation, and $\Delta$ 
effects that are not taken into account in 
these standard calculations.
 
\begin{figure}
\includegraphics[width=0.75\linewidth,clip]{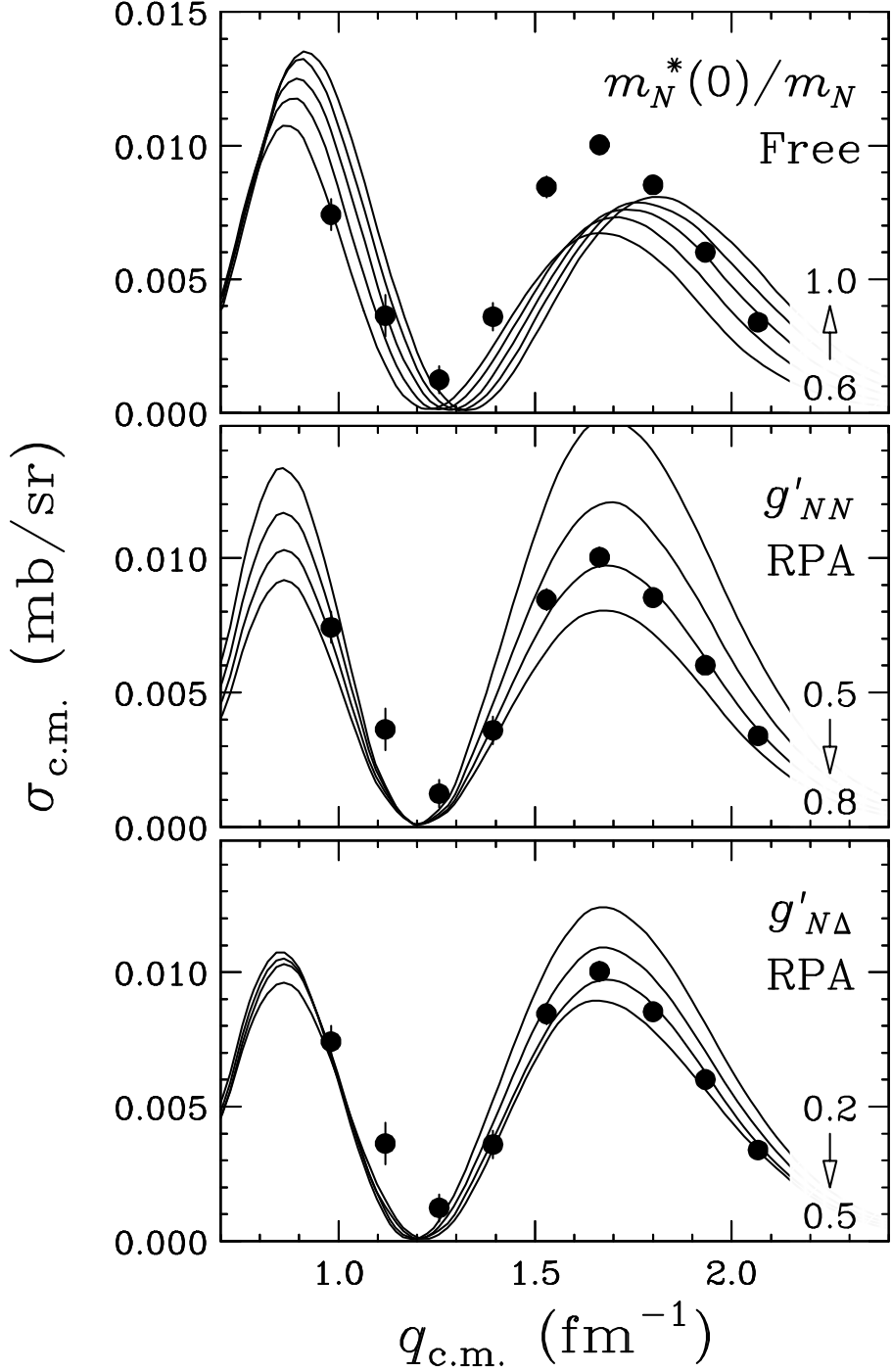}%
\caption{
 The top panel shows the $m^*(0)/m_N$ dependence of the 
DWIA calculations with the free response function.
 The middle panel represents the $g'_{NN}$ dependence 
of the calculations employing the RPA response function 
with fixed $g'_{N\Delta}$ = 0.4 and $m^*(0)/m_N$ = 0.7.
 The bottom  panel denotes the $g'_{N\Delta}$ dependence 
of the calculations employing the RPA response function 
with fixed $g'_{NN}$ = 0.7 and $m^*(0)/m_N$ = 0.7.
 The curves in each panel represent calculations where 
the parameters in the ranges shown are changed in steps 
of 0.1.
\label{fig:rpa}}
\end{figure}

 The non-locality of the nuclear mean field can be 
included by introducing a local effective mass
approximation in the form of 
\begin{equation}
m^*(r) = m_N-\frac{f_{\rm WS}(r)}{f_{\rm WS}(0)}(m_N-m^*(0)),
\label{eq:effmass}
\end{equation}
where $m_N$ is the nucleon mass and $f_{\rm WS}(r)$ is a 
WS radial form.
 The upper panel of Fig.~\ref{fig:rpa} shows the $m^*$ 
dependence of the DWIA calculations with the free response function 
that were performed using the computer code {\sc crdw} 
developed by the Ichimura group \cite{prc_63_044609_2001}  
for the analysis of QES data. 
 The $0^-$ component of the free response is 
configured as a pure $0p_{1/2}^{-1}1s_{1/2}$ 
transition.
 The DWIA result with $m^*(0)$ = $m_N$ 
is in good agreement with the calculation 
employing the corresponding SM wave function 
represented by the dashed curve in 
Fig.~\ref{fig:xsec}.
 Thus we have applied the same normalization factor 
of 0.64 to the results shown in Fig.~\ref{fig:rpa}.
 The angular distribution shifts to 
lower $q_{\rm c.m.}$ when decreasing $m^*(0)$.
 As seen in the upper panel of Fig.~\ref{fig:rpa}, 
a value of $m^*(0)/m_N$ $\simeq$ 0.7 improves the 
agreement with the data, 
consistent with theoretical estimations 
\cite{plb_126_421_1983,npa_481_381_1988}.
 However there is still a large discrepancy between experimental 
and theoretical results around $q_{\rm c.m.}$ $\simeq$ 
$1.7\,\mathrm{fm^{-1}}$.

 Next, we discuss the RPA correlation and $\Delta$ effects.
 We performed DWIA calculations with the RPA
response functions employing
the $\pi+\rho+g'$ model interaction $V_{\rm eff}$ and
the meson parameters from a Bonn potential which  
treats the $\Delta$ explicitly \cite{pr_149_1_1987}.
 The $V_{\rm eff}$ is the sum of the one-$\pi$ and
one-$\rho$ exchange interactions, and the Landau-Migdal 
(LM) interaction $V_{\rm LM}$ specified by the LM
parameters, $g'_{NN}$, $g'_{N\Delta}$, and $g'_{\Delta\Delta}$, as
\begin{equation}
\begin{split}
&V_{\rm LM} 
=\left[\frac{f_{\pi NN}^2}{m_{\pi}^2}g'_{NN} 
(\bm{\tau}_1\cdot\bm{\tau}_2)(\bm{\sigma}_1\cdot\bm{\sigma}_2) \right.\\
&+ \frac{f_{\pi NN}f_{\pi N\Delta}}{m_{\pi}^2}g'_{N\Delta} 
\left\{
\left((\bm{\tau}_1\cdot\bm{T}_2)(\bm{\sigma}_1\cdot\bm{S}_2) 
+ {\rm h.c}. \right)  + (1\leftrightarrow 2)
\right\}\\
&+ \left. \frac{f_{\pi N\Delta}^2}{m_{\pi}^2}g'_{\Delta\Delta}
\left\{(\bm{T}_1\cdot\bm{T}^{\dag}_2)(\bm{S}_1\cdot\bm{S}^{\dag}_2) 
+ (1\leftrightarrow 2) \right\}\right] \delta(\bm{r}_1 - \bm{r}_2),
\end{split}
\end{equation}
where $\bm{\sigma}$ ($\bm{\tau}$) is the nucleon Pauli spin (isospin) 
matrix, $\bm{S}$ ($\bm{T}$) is the spin (isospin) transition operator 
that excites {\it N} to $\Delta$,  
$f_{\pi NN}$ ($f_{\pi N\Delta}$) is the $\pi NN$ ($\pi N\Delta$) 
coupling constant, and $m_{\pi}$ is the pion mass.
 The middle panel of Fig.~\ref{fig:rpa} 
shows the $g'_{NN}$ dependence for 
$g'_{NN}$ = 0.5--0.8 in 0.1 steps with the fixed 
$g'_{N\Delta}$ = 0.4 and $m^*(0)/m_N$ = 0.7.
 The lower panel shows the $g'_{N\Delta}$ dependence for 
$g'_{N\Delta}$ = 0.2--0.5 in 0.1 steps with the fixed 
$g'_{NN}$ = 0.7 and $m^*(0)/m_N$ = 0.7.
 We fixed $g'_{\Delta\Delta}$ = 0.5 \cite{prc_23_1154_1981} 
since the dependence of the calculated results on this 
parameter is very weak.
 The calculated angular distributions depend strongly on 
$g'_{NN}$ in the whole $q_{\rm c.m.}$ range, 
whereas a strong $g'_{N\Delta}$ dependence is observed only 
around $q_{\rm c.m.}$ $\simeq$ $1.7\,\mathrm{fm^{-1}}$.
 The most probable choices are $g'_{NN}$ $\simeq$ 0.7 and 
$g'_{N\Delta}$ = 0.3--0.4.
 As a result, the $V_{\rm eff}$ in the {\it NN} channel
are close to zero whereas that in the $N\Delta$ channel
becomes very attractive \cite{prc_69_054609_2004}.
 This attraction causes the pionic enhancement.

 In conclusion, our high-resolution measurement of 
${}^{16}{\rm O}(p,p'){}^{16}{\rm O}(0^-,T=1)$ has 
enabled us to search for a pionic enhancement at an intermediate 
energy of $T_p$ = 295 MeV where the theoretical DWIA calculations 
are reliable owing to the simple reaction mechanism.
 A significant enhancement has been observed 
around $q_{\rm c.m.}$ $\simeq$ 
$1.7\,\mathrm{fm^{-1}}$ compared to standard DWIA calculations 
with SM wave functions.
 The DWIA analyses employing RPA response functions 
with the $\Delta$ isobar
allowed the determination of $m^*(0)/m_N$ $\simeq$ 0.7 and 
a set of the Landau-Migdal parameters of 
$g'_{NN}$ $\simeq$ 0.7 and $g'_{N\Delta}$ = 0.3--0.4, 
in good agreement with estimates based on
other experimental data \cite{nucl_ex_0411055}.
 The present direct measurement strongly indicates the existence of 
the pionic enhancement in nuclei that may be attributed to a 
precursor phenomena of the pion condensation.

 We thank the RCNP cyclotron crew for providing a good 
quality beam.
 This work was supported in part by the Grants-in-Aid for
Scientific Research Nos.~12740151, and 14702005 
of the Ministry of Education, Culture, Sports, 
Science, and Technology of Japan.


%

\clearpage

\clearpage

\clearpage

\clearpage

\clearpage

\bibliography{wakasa}

\end{document}